\documentclass[aps,pra,epsf,superscriptaddress,amsmath,amssymb,amsfonts,twocolumn,showpacs,nofootinbib]{revtex4-1}

\usepackage{graphicx}
\usepackage{epsfig}
\usepackage{dcolumn}
\usepackage{bm}
\usepackage{braket}
\usepackage{amsmath}
\usepackage{mathtools}
\usepackage{graphicx,color,xcolor}
\usepackage{hyperref}
\newcommand{\abs}[1]{\left| #1 \right|} 
\usepackage{hyperref}
\usepackage{multirow}
\usepackage{afterpage}
 \usepackage{overpic} 
 \usepackage{float}
 \usepackage{physics}

\begin{document}

\title{Restoring the topological edge states in a finite optical superlattice}

\author{ A. Katsaris$^*$}
\affiliation{Department of Physics, National and Kapodistrian University of Athens, GR-15784 Athens, Greece} 
\author{ I. A. Englezos$^*$}
\affiliation{Zentrum für Optische Quantentechnologien, Luruper Chaussee 149, Universität Hamburg, 22761 Hamburg, Germany}
\author{ C. Weitenberg}
\affiliation{Institut für Quantenphysik,  Luruper Chaussee 149, Universität Hamburg, 22761 Hamburg, Germany} 
\affiliation{The Hamburg Centre for Ultrafast Imaging, University of Hamburg, Luruper Chaussee 149, 22761 Hamburg, Germany} 
\author{F. K. Diakonos}
\affiliation{Department of Physics, National and Kapodistrian University of Athens, GR-15784 Athens, Greece} 
\author{P. Schmelcher}
\affiliation{Zentrum für Optische Quantentechnologien,  Luruper Chaussee 149, Universität Hamburg, 22761 Hamburg, Germany}
\affiliation{The Hamburg Centre for Ultrafast Imaging,
University of Hamburg, Luruper Chaussee 149, 22761 Hamburg, Germany}

\date{\today}

\begin{abstract}
We consider the emergence of edge states in a finite optical lattice and show that the boundaries of the lattice play a decisive role for their location in the corresponding energy spectrum. We introduce a simple parametrisation of the boundaries of the optical lattice and demonstrate the existence of an optimal choice of the values of the parameters which lead to an approximate restoration of chiral symmetry. A crucial property of this optimization is the suppression of tunneling between next-nearest neighboring wells of the lattice. This in turn allows the mapping of the optical lattice set-up to a finite SSH model. The topological character of the emerging edge states is discussed.
\end{abstract}
 
\maketitle
\def\thefootnote{*}\footnotetext{These authors contributed equally.}

\section{Introduction} 

Ever since the experimental realization of Bose-Einstein condensation (BEC)~\cite{BECexpAnderson,BECexpBradley,BECexpDavis} ultracold atomic platforms have emerged as a highly versatile test-bed for a wide range of phenomena in atomic, molecular and condensed matter systems~\cite{BlochLatticeRev,BlochNature2012}. A crucial step in that direction is the ability to precisely control the inter-atomic interaction strengths, via Feshbach resonances~\cite{FeshbachExpCourteille,FeshbachExpInouye,Chin2010RevModPhysFeshbach}. Modern optics can be utilized so that the shape of the external confinement results in almost arbitrary trap geometries, including setups of low-dimensional lattices~\cite{bakr2009quantum}. This level of control has enabled among others the realization of Hubbard models and the study of the Superfluid-Mott insulator transition~\cite{JakschBH,GreinerExplattice}. Moreover, it has facilitated the realization of topological phases of matter related to the quantum Hall effect and topological insulators such as the Su-Schrieffer-Heeger (SSH) model~\cite{DalibardReview, weitenberg2021tailoring}.

The simplicity of the SSH model establishes it as an ideal starting point for understanding the topological phases of matter.
One of the most important features of the SSH model is its chiral symmetry, which is strongly connected to its topological properties. Namely, since there is a strong bulk-edge correspondence (BDI symmetry class \cite{Ryu_2010}), one can predict when topologically protected edge states will be supported by the system. Those states are of great interest because of their robustness against certain disorders, rendering them promising candidates for quantum information processing \cite{PhysRevA.103.052409,PhysRevResearch.2.033475,PhysRevA.106.022419}. The SSH model can be easily extended to describe more complex systems, such as higher dimensional systems with enriched topological phases \cite{PhysRevB.106.245109} or including interactions \cite{PhysRevLett.110.260405, PhysRevA.99.053614,PhysRevB.107.054105}.       

Realizations of the SSH model in optical lattices are mainly focused on the study of the quantized charge transport both for fermionic \cite{Nakajima_2016} and bosonic \cite{Lohse_2015} atoms. In this context, the parameters of the discrete SSH model are derived from the continuous optical lattice potential, by fitting their corresponding Bloch spectra primarily in the two lowest bands, whose gap determines the topological properties of the system~\cite{DalibardReview}. Importantly,  in principle next-nearest neighbor (NNN) hopping is always present in optical lattices, although it is significantly suppressed for deep potential wells and it is usually ignored in studies of the SSH model. However, from the theoretical side, even a small next-nearest neighbor hopping amplitude explicitly breaks the underlying chiral symmetry of the SSH model \cite{PhysRevB.89.085111,PhysRevB.99.035146}, leading to an extended SSH model (eSSH) belonging to a different symmetry class (AI) and possessing a different topological invariant. Additionally, edge states in 2D topological systems were observed only recently by adding a sharp wall potential to a driven honeycomb lattice \cite{braun2023realspace} and a rotating trap \cite{yao2023observation}. Other observations of edge states relied on artificial dimensions using internal states for defining sharp edges \ \cite{PhysRevLett.108.133001, Mancini_2015, Stuhl_2015, Kanungo_2022}. Edge states were also probed in tweezer arrays with Rydberg interactions \cite{doi:10.1126/science.aav9105}, where the edges also occur naturally.

For the above reasons, we focus on a systematic study of the conditions under which a finite continuous optical lattice may be mapped, within the tight-binding approximation, to an SSH model. Specifically, we demonstrate and analyze the problems that emerge when finite confined systems are considered, and we provide the essential tools to control them. In particular, we introduce a simple extension of the potential domain of the optical lattice, and we find the optimal choice of its parameters values so that the effects of the confinement are maximally suppressed. Finally, we address the presence of NNN hopping when the potential wells are not deep enough, and propose criteria so that the corresponding terms can be omitted.   

This work is structured in the following way. In Section~\ref{sec:setup} we mention the basic characteristics of the SSH and eSSH models and describe the optical superlattice potential we employ to realise them. The explicit mapping from the continuous to the discrete system is also presented. 
Then, in Section~\ref{sec:Continuousresults} we analyze the criteria and in particular boundary conditions under which topological edge states emerge in the finite continuous system, via directly solving the corresponding Schr\"odinger equation.
In Section~\ref{sec:ssh-Results} we present the results derived from the mapping to the discrete (tight-binding) system and examine their deviations from the (extended) SSH model with the same bulk parameters. 
Finally, in Section~\ref{sec:SummaryAndOutlook} we summarise our results and highlight future perspectives.

\section{Tight-binding models and Optical lattice Potentials}\label{sec:setup} 

In this section, we begin by noting the basic characteristics of the SSH model and its extension (eSSH) when NNN hopping terms are considered (Sec.~\ref{sec:ssh_model}). We then proceed with a continuous setup involving a superlattice optical potential, which enables the implementation of the SSH model in ultracold atomic platforms (Sec.~\ref{sec:SL}). Lastly, we present the tight binding (TB) approximation, which we employ for the detailed comparison between the continuous and the corresponding discrete system (Sec.~\ref{sec:TB}).

\subsection{SSH and eSSH models}\label{sec:ssh_model}

\begin{figure}[ht]
    \begin{center}
            \includegraphics[width = 0.48\textwidth]{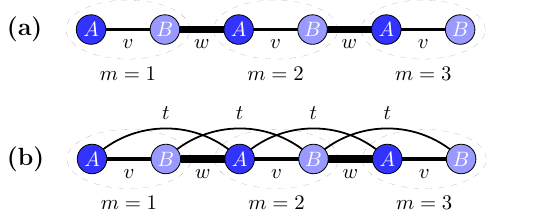}
    \end{center}
    \caption{Schematics of (a) SSH and (b) eSSH models with $M=3$ cells.}
    \label{SSH_schematic}
\end{figure}

The Hamiltonian of an SSH model \cite{PhysRevLett.42.1698,Coffee&Donuts} with $M$ unit cells can be written as:

\begin{equation}
\mathcal{H}_\text{SSH} = v\sum_{m=1}^{M}a^\dagger_{m}b_{m} + w\sum_{m=1}^{M}a^\dagger_{m+1}b_{m} + h.c
\end{equation}
where $v,w$ are the intracell and intercell hopping amplitudes, $\hat{a}^\dagger_m$ ($\hat{a}_m$) and $\hat{b}^\dagger_m$ ($\hat{b}_m$) are the creation (annihilation) operators, creating (annihilating) a particle in the $m$ cell on the $A/B$ sublattice respectively. A schematic of the SSH model is provided in Fig.~\ref{SSH_schematic}(a). We also consider an extension of the SSH model with the addition of NNN hopping terms of the form:

\begin{equation}
\mathcal{H}_\text{NNN} = t\sum_{m=1}^{M-1}\left(\hat{a}^{\dagger}_{m+1}\hat{a}_{m}+\hat{b}^\dagger_{m+1}\hat{b}_{m}\right) + h.c.
\end{equation}
where $t$ is the NNN hopping amplitude (see Fig.~\ref{SSH_schematic}(b)). The Hamiltonian of this extended SSH (eSSH) is:
\begin{equation}
\mathcal{H}_\text{eSSH} = \mathcal{H}_\text{SSH} + \mathcal{H}_\text{NNN}
\label{eSSH_eq}
\end{equation}

\noindent These two models are of great interest because they can both support topologically protected edge states (TESs), i.e. states that are localized on the edges of the lattice and their energies reside in the center of the band gap. We emphasize that in the SSH model the existence and protection of TESs are connected to the chiral symmetry of the system (BDI symmetry class) \cite{Ryu_2010}, while in the case of the eSSH model they are related to the inversion symmetry (AI symmetry class) \cite{PhysRevB.89.085111,PhysRevB.99.035146}. Specifically, in the SSH model when $v>w$ the system is in the topologically trivial phase and it does not support edge states. In contrast, for $v<w$ the system resides in the topologically non-trivial (i.e. topological) phase where it supports two edge states. On the other hand, for the case of eSSH, even though the bulk-edge correspondence holds, specifics of the edge states, such as their position on the energy spectrum or their robustness, can not be easily predicted. However, when edge states are supported by eSSH, they are in fact of topological nature. In most of the cases considered below the parameter $t$ will always be relatively small in comparison to the values of the $v$ and $w$ parameters, hence the NNN terms could be treated as a perturbation.

\subsection{Superlattice potential}\label{sec:SL}

An optical lattice can be implemented experimentally by forming a standing wave, utilizing two counter propagating laser beams. A sequence of $M$ double wells as required to realize the SSH and eSSH models, can be achieved by superimposing two such standing waves with different frequencies~\cite{Anderlini_2007, F_lling_2007, DalibardReview}, leading to the superlattice (SL) potential:
\begin{align}\label{SL_Potential}
V_{SL}(x) = V_s\cos^2{\left(2k_{r} x\right)} + V_l\cos^2{\left(k_{r} x \right)}
\end{align}
where $V_s,~V_l$ are the amplitudes of the two standing waves, $k_{r}= 2\pi/\lambda_0$ is the single-photon recoil momentum, and $\lambda_0$ is the wavelength of the lattice. For a system of $M$ cells and total length $2L$ it holds that $\lambda_0 = 2L/M$.

\begin{figure}[ht]
\begin{center}
\includegraphics[scale=0.42]{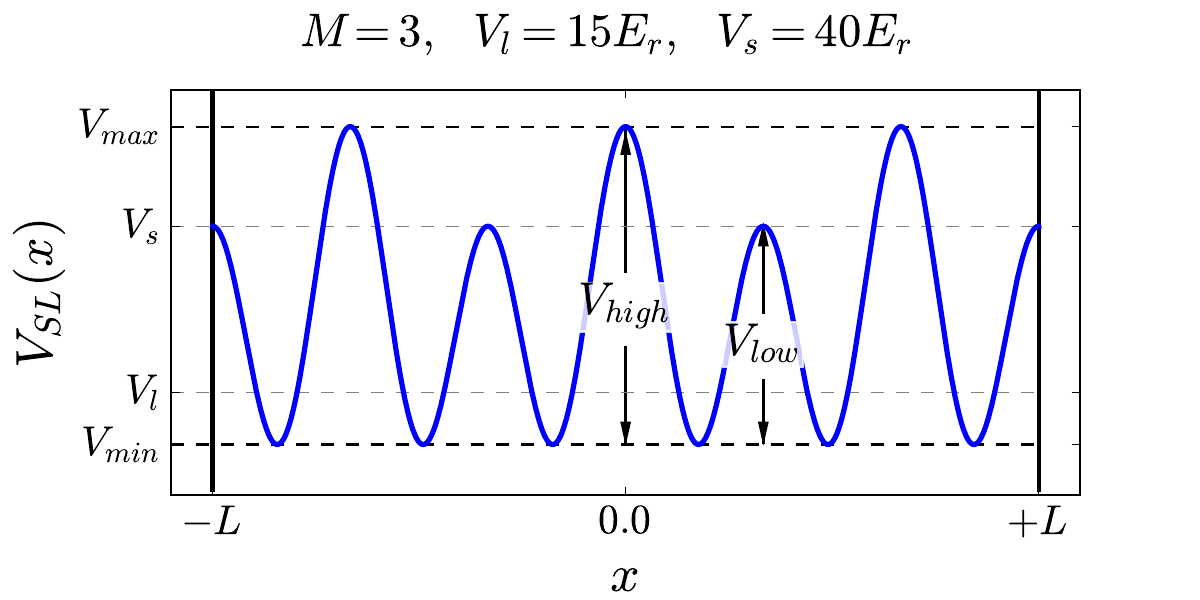}
    \caption{Schematic of the SL potential for $M=3$, $V_l = 15E_r$ and $V_s = 40E_r$, and $x$ in units of $k_{r}^{-1}$.}
    \label{SL_schematic}
\end{center}
\end{figure}

The stationary Schr\"odinger equation
\begin{equation}
\hat{H}\ket{\Psi_n} = E_n\ket{\Psi_n},~\text{with}~ \label{Schrodinger}~\hat{H} = -\frac{\hslash^2}{2m} \frac{\partial^2}{\partial x^2} + \hat{V_{SL}}  
\end{equation}
can then be solved numerically, e.g. via exact diagonalization (ED), to obtain the complete spectrum of eigenstates and eigenenergies of the system. In the following, our analysis will focus on the lowest band, i.e. the eigenstates with the $2M$ lowest laying eigenvalues.

The heights of the two barriers of the SL potential shown in Fig. \ref{SL_schematic} are given by the following expressions:
\begin{align}
V_{low}  = V_s(1 - \frac{V_l}{4V_s})^2,\quad V_{high} &= V_s(1 + \frac{V_l}{4V_s})^2
\end{align}
Evidently, changing one of the amplitudes $V_{s,l}$ results in a non-linear shift of both the relative and absolute heights of the potential barriers. 
Hence, it is convenient to express our results in terms of the height of the higher barrier $V_{high}$ and the ratio between the lower and higher barrier $u = V_{low}/V_{high}$. Finally, for computational convenience, we recast the Schr\"odinger equation in a dimensionless form by expressing the length in terms of $k_r^{-1}$ and the energy in terms of the recoil energy $E_r = \hbar^2 k^2_{r}/2m$, where $m$ is the atomic mass and $\hbar$ the reduced Plank constant.

\subsection{Tight Binding Approximation}\label{sec:TB}

In the context of the tight binding (TB) approximation, we assume that in a deep enough lattice, the description of the system can be accurately truncated to the first, or first few, energy bands~\cite{WannierReview,tweezerArrays2024}. For a system of $M$ double wells (cells), i.e. a system with $\mathcal{N}=2M$ minima (sites), restricted to the lowest band, the TB Hamiltonian takes the form:
\begin{equation}\label{TB_Hamilt}
    \hat{\mathcal{H}}_{TB} = \sum_{i=1}^{\mathcal{N}}\sum_{j=1}^{\mathcal{N}} h_{i,j} \hat{\alpha}^\dagger_{i}\hat{\alpha}_{j},
\end{equation}
where $\hat{\alpha}^\dagger_{i}$ ($ \hat{\alpha}_{i}$) are the creation (annihilation) operators, creating (annihilating) either a bosonic or a fermionic particle in the lowest band at the site $i$, and the matrix elements
\begin{equation}\label{TB_elements}
    h_{i,j} = \int w^*_i(x) \Big[ \frac{\hbar^2}{2m}\frac{\partial^2}{\partial x^2} - \hat{V_{SL}} \Big] w_j(x) dx
\end{equation}
are defined in terms of the Wannier functions $w_i(x)$ localized at each site $i$. 
In the following, we define the Wannier functions as
\begin{equation}
    w_i(x)=\bra{x}\ket{\chi_i}
\end{equation}
where $\ket{\chi_i}$ are the eigenstates of the position operator restricted to the lowest band ($\hat{\mathcal{X}}_{band}$), which can be obtained by solving the eigenvalue problem $\hat{\mathcal{X}}_{band} \ket{\chi_i} = \chi_i  \ket{\chi_i}$, with 
\begin{equation}
\hat{\mathcal{X}}_{band} =  \sum_{n=1}^{\mathcal{N}}  \sum_{m=1}^{\mathcal{N}} \ket{\Psi_n}\bra{\Psi_n}\hat{x}\ket{\Psi_m}\bra{\Psi_m},
\end{equation}
and $\ket{\Psi_{n}}$, $\ket{\Psi_{m}}$ fulfilling the eigenvalue problem given by Eq. \eqref{Schrodinger}.
In 1D this definition of the Wannier functions has been shown to produce uniquely defined maximally-localized Wannier functions, even when generalized to take into account higher bands~\cite{WannierStates}. In the following, we aim to determine the conditions under which the TB Hamiltonian~\eqref{TB_Hamilt}, corresponding to the finite continuous system subjected to the SL potential, 
can be accurately mapped to the SSH or eSSH model.

\begin{figure*}[ht]
\begin{center}
\begin{overpic}[width=0.96\textwidth]{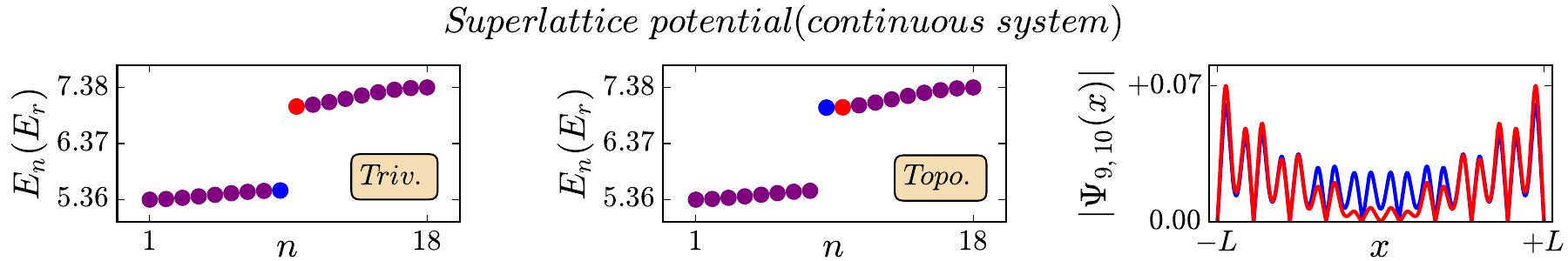}
\put(9,10) {\textbf{(a)}}
\put(43.5,10) {\textbf{(b)}}
\put(79,10) {\textbf{(c)}}
\end{overpic}
\\[10pt]
\begin{overpic}[width=0.96\textwidth]{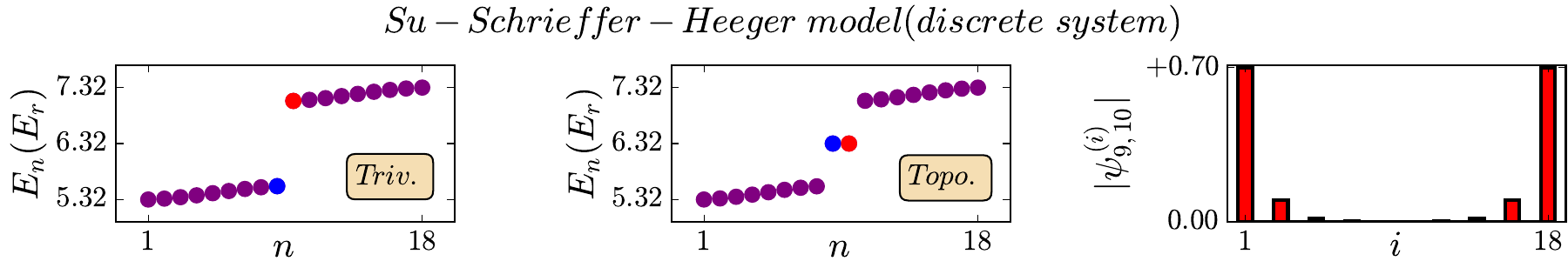}
\put(9,10) {\textbf{(d)}}
\put(43.5,10) {\textbf{(e)}}
\put(81,10) {\textbf{(f)}}
\end{overpic}
\end{center}
\caption{(a,b) The energy eigenvalues of the superlattice potential with hard-wall boundary conditions for $V_{high}=40E_R$, $V_{low}=20E_R$, and $M=9$ in the trivial and topological regime obtain via ED. (c) The non-topologically protected edge states appearing in the presence of hard-wall boundary conditions for the continuous system ($x$ in units of $k_{r}^{-1}$). (d,e) The energy eigenvalues of the SSH-model in the trivial and the topological regime.
(f) The edge states of the associated SSH-model. The relevant hopping amplitudes for the SSH model, in (d)-(f), have been obtained from the bulk values of the TB Hamiltonian i.e. $v,w = h_{M,M \pm 1}$ and $\mu = h_{M,M}$. The overall energy scale offset of the graphs (a,b) and (d,e) is related to the larger on-site amplitudes of the edge sites in comparison to the bulk sites, which in this case we have not taken into account.}
\label{fig:HardWallCase}
\end{figure*}
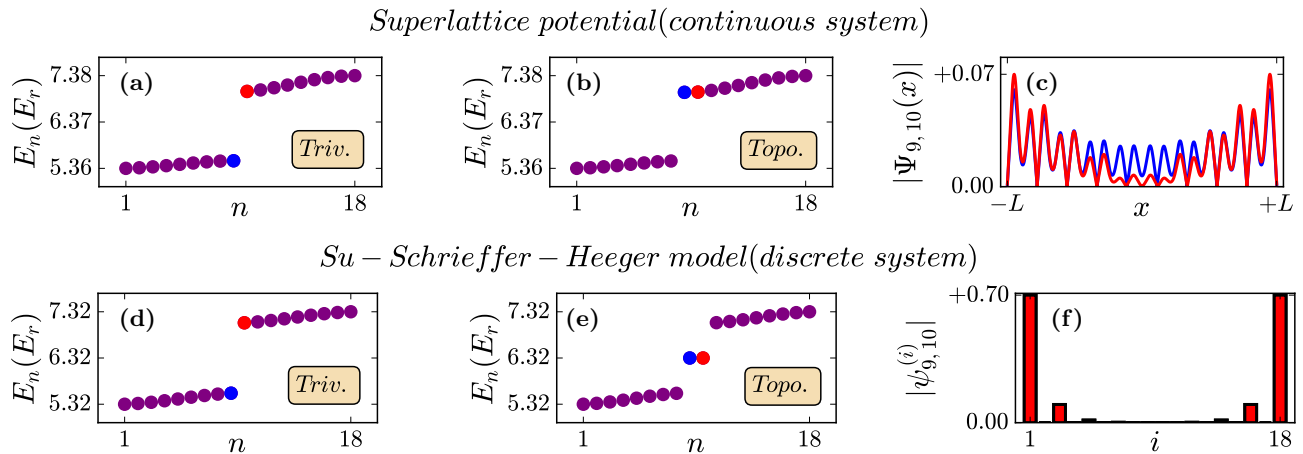

\section{Impact of the Boundary Conditions on the finite continuous system}\label{sec:Continuousresults} 

In order to obtain a uniquely defined set of eigenstates and eigenergies (up to an overall phase), specific boundary conditions have to be implemented when we solve the Schr\"odinger equation. For a periodic system, e.g. in a ring geometry, one has to employ periodic boundary conditions (PBC). In this case, it has been shown that the lowest band of the energy spectrum of the continuous system subjected to the SL potential given by Eq~\eqref{SL_Potential}, is in good agreement with that of the corresponding discrete SSH model with PBC~\cite{DalibardReview,GreinerExplattice}.
However, in the case of PBC the system cannot exhibit edge states by construction. Hence, in order to directly observe topological edges states we have to consider finite lattices with open boundary conditions, i.e. systems with clearly defined edges.

When considering finite systems, the choice of boundary conditions is not unique.
A common choice, is the consideration of hard wall boundary conditions (HWBC), i.e. to demand that the eigenstates are exactly zero (vanish) at the boundaries of the potential landscape.
Moreover, we expect a vanishing probability of observing a particle in the regions where the energy lies below the potential strength, namely in the classically forbidden regions. So, the energetically low-lying eigenstates are expected to take values close to zero in the vicinity of the positions of the potential's local and global maxima. In the context of optical lattices the boundary conditions are usually considered at its maxima, so the eigenstates are expected to not exhibit an abrupt transition of their profiles, even in the case of HWBC. Ultimately, from a theoretical point of view HWBC seems at first to be a particularly natural choice for our system, especially when the description is restricted to the lowest band. However, in this section we illustrate that the they are not an appropriate choice for capturing topologically protected edge states in the continuous finite system (Sec.~\ref{sec:HWBC}) and we propose an alternative boundary which instead allows for their emergence (Sec.~\ref{sec:Extention}).

\subsection{The case of hard wall boundary conditions}\label{sec:HWBC}

\begin{figure*}[ht]
\begin{center}
\begin{overpic}[width=0.94\textwidth]{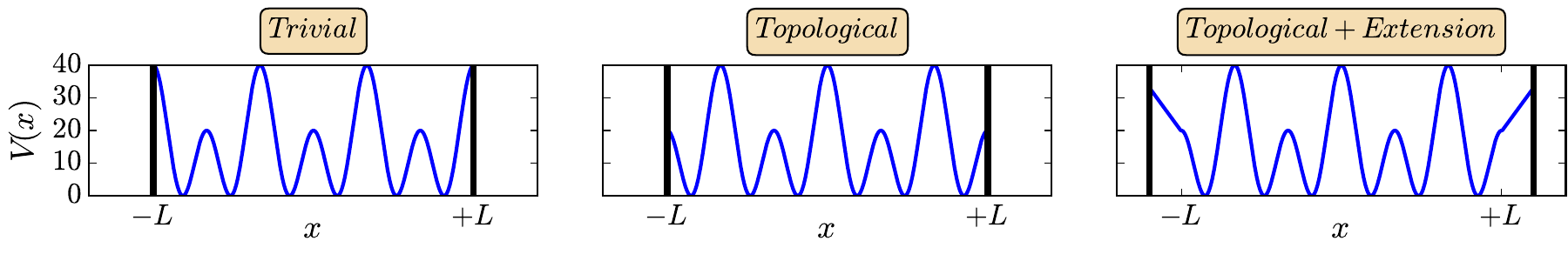}
\put(12.5,14.25) {\textbf{(a)}}
\put(43.5,14.25) {\textbf{(b)}}
\put(71,14.25) {\textbf{(c)}}
\end{overpic}
\\[10pt]
\begin{overpic}[width=0.40\textwidth]{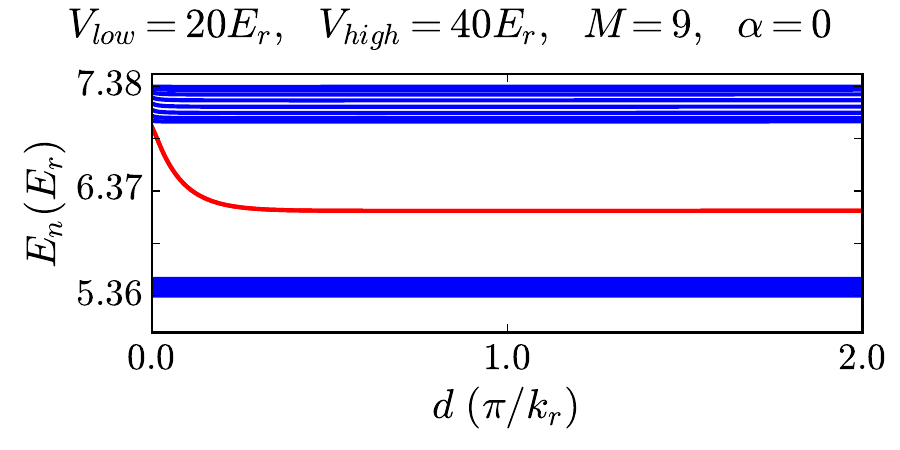}
\put(87,30) {\textbf{(d)}}
\end{overpic}
\quad\qquad\quad
\begin{overpic}[width=0.40\textwidth]{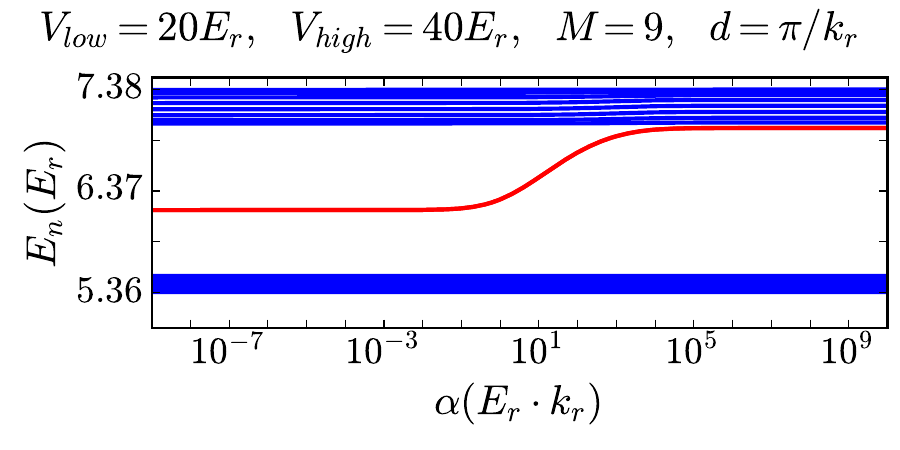}
\put(89,30) {\textbf{(e)}}
\end{overpic}
\end{center}
\caption{The superlattice potential with hard-wall boundary conditions for $V_{high}=40E_r$, $V_{low}=20E_r$, and $M=3$ in the topologically (a) trivial, (b) non-trivial regime. (c) Same as (b) with the addition of the linearly extended boundary for a slope $\alpha=7 Er\cdot k_r$ and $d = 0.3L_{cell}$. In (a)-(c) $V(x)$ is in units of $E_r$ and $x$ in units of $k_r^{-1}$. (d) The energy eigenvalues for $M=9$ for increasing the length of the extended boundary $d$ and vanishing slope ($\alpha=0$). (e) Same as (d), but for increasing slope ($\alpha$) and fixed length $d=\pi/k_r$.}
\label{fig:LinearExtension}
\end{figure*}

First, we begin our analysis with a topologically trivial setting where no edge states emerge. Namely, we consider the continuous system with HWBC, potential barriers of heights $V_{high}=40E_r$, $V_{low}=20E_r$ and $M=9$ cells and compare the energy spectrum in the lowest band with the one obtained from the SSH model with open boundary conditions (OBC) as shown in Fig.~\ref{fig:HardWallCase}(a) and (d) (see also Fig.~\ref{fig:LinearExtension}(a) for a schematic of the potential in the topologically trivial regime). Evidently, we indeed find a good agreement with the corresponding SSH model with OBC in the topologically trivial regime.

Surprisingly, the situation is profoundly different when considering the topological phase for the same parameters and HWBC (see Fig.~\ref{fig:LinearExtension}(b) for a schematic of the potential). Specifically, the energy spectrum for the continuous system in Fig.~\ref{fig:HardWallCase}(b) exhibits a structure featuring one sub-band with $M-1$ states and a higher sub-band with $M+1$ states, in sharp contrast with the spectrum of the corresponding SSH model depicted in Fig.~\ref{fig:HardWallCase}(e), which features two sub-bands containing $M-1$ states and the two edge states residing in the center of the gap. Interestingly, the states in the middle of the spectra of both the continuous system and the SSH model ($\Psi_{M}$, $\Psi_{M+1}$), exhibit localization at the edges as shown in Fig.~\ref{fig:HardWallCase}(c) and (f).

However, the edge localized states predicted by the continuous system have support in both odd and even sites --unlike the topologically protected states of the SSH model-- which is a strong signature of the particle-hole symmetry breaking of the system. Moreover, we note that the edge localized states of the continuous system are significantly less localized, and become well separated from the bulk only in the case of very large or deep lattices (see Appendix \ref{app:LargeSystem} for an illustrative example).

Finally, it is clear that the edge localized states in the continuous system are not topologically protected, since they are not separated by a sub-band gap from the bulk states. We interpret the results presented here as an indication that the HWBC cause an offset to the potential experienced by an atom at the edge sites $1$ or $2M$ as compared to an atom residing in the center of the lattice. Evidently, this seems irrelevant when considering a topologically trivial lattice, however it has a profound effect in a topological lattice where edge states may be found.

\subsection{Extension of the potential domain}\label{sec:Extention}
\begin{figure*}[ht]
\begin{center}
\begin{overpic}[width=0.96\textwidth]{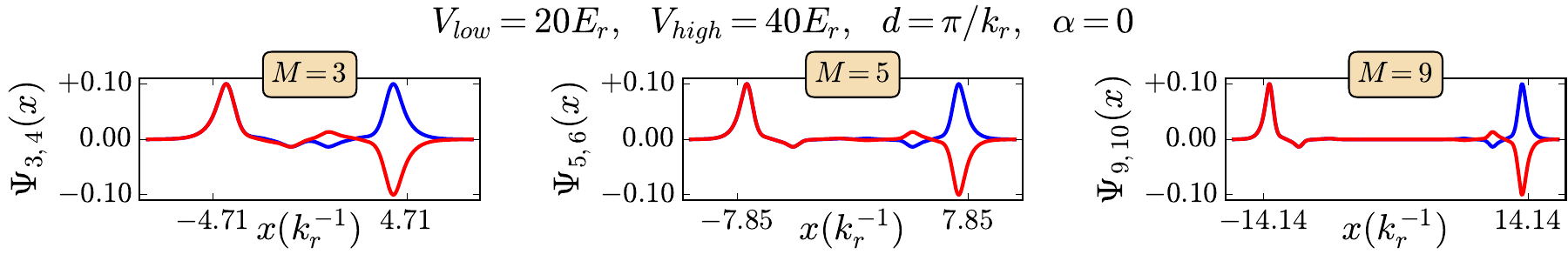}
\end{overpic}
\end{center}
\caption{The topologically protected edge states appearing in the presence of the linearly extended boundary at $d=\pi/k_r$ and $\alpha=0$, for systems with $M=5,9,13$.}
\label{fig:LinearExtensionEdge}
\end{figure*}

To address the discrepancy between the continuous system and the SSH model we introduce an extension to the SL potential, in order to simulate a more physical boundary than the infinite wall. This extension should be simple enough to aid the theoretical analysis and at the same time to be experimentally feasible. So we consider a linear extension of length $d$ and slope $\alpha$, as follows (see also Fig.~\ref{fig:LinearExtension}(c)):

\begin{equation}
\label{Potential_Extension}
V_{ext}(x)= \begin{cases}
      ~V_s - \alpha(x+L),  & - (L + d)\leq x\leq -L\\[5pt]
      ~0,& -L<0< L \\[5pt]
      ~V_s + \alpha(x-L), & L\leq x\leq L+d
      \end{cases}
\end{equation}
Quite surprisingly, this rather simple extension readily fixes the discrepancy of the spectral behaviour between the continuous system and the SSH model. As shown in Fig.~\ref{fig:LinearExtension}(d) upon increasing the length $d$ of the linear extension (for fixed $\alpha=0$) the edge state energies move towards the center of the gap between the two sub-bands. In contrast, upon increasing the slope $\alpha$ (for fixed $d=\pi/k_r$), and hence recovering the effect of a sharp wall at the edge (in the limit $\alpha\rightarrow \infty$ and $d \neq 0$), progressively the edge state energies shift towards the higher sub-band. This is shown in Fig.~\ref{fig:LinearExtension}(e), where the logarithmic scale of the $x$-axis highlights the relatively large values of the slope $\alpha$ required for the HWBC effect to re-emerge. Moreover, we indeed observe topologically protected edge states, even for smaller systems with $M=3,5$ and $M=9$ as shown in Fig.~\ref{fig:LinearExtensionEdge}.

Finally, in Fig.~\ref{fig:LinearExtension}(d) and (e) we observe asymptotically constant behaviour of the eigenenergies when $d$ is increased (or $\alpha$ decreased) beyond a certain value. Based on this behavior we can make an estimation about the required length of the linearly extended boundary, so that the energies of the two edge states reside in the center of the two sub-bands. For all values of the parameters ($M$, $V_{high}$ and $V_{low}$), we find that a relatively small length of the boundary extension is needed $d\leq 0.5\pi/k_r$ for the energies of the edges states to saturate with respect to $d$, indicating that only the local behavior of the boundary extension around the first and the last site is relevant. 

\section{Tight-Binding Approximation Analysis}\label{sec:ssh-Results}

In this section we are focusing on the results related to the corresponding discrete system, i.e. the form of the tight-binding Hamiltonian ($\hat{\mathcal{H}}_{TB}$) and the solution of its eigenvalue problem. Specifically, we introduce a decomposition of the $\hat{\mathcal{H}}_{TB}$ into two terms, the $\hat{\mathcal{H}}_{0}$ for the bulk values and the $\delta \hat{\mathcal{H}}$ that expresses the deviations due to the HWBC (Sec. \ref{delta_H}). Then, we analyse the behaviour of $\delta \hat{\mathcal{H}}$ with respect to the parameters of the extension of the potential ($d$ and $\alpha$), and we find the region in which it is minimised (Sec. \ref{delta_H_zero}). Finally, we consider systems with $\delta\hat{\mathcal{H}}=0$ and we focus on the behaviour of $\hat{\mathcal{H}}_{0}$ (Sec. \ref{NNN_term}). 

\subsection{Introducing the $\vb*{\mathcal{H}_0}$ and $\vb*{\delta \mathcal{H}}$ terms}
\label{delta_H}
In order to make our analysis simpler, we establish specific notations for the hopping amplitudes $h_{i,j}$, where $i$ and $j$ are the indices of the sites of the discrete system (corresponding to the minima of the superlattice potential). So we set the on-site ($i=j$) amplitudes as $\mu_i$, the nearest neighbour ($\abs{i-j}=1$) hopping amplitudes as $J_{v_i}$ and $J_{w_i}$ and the next-nearest neighbour hopping amplitudes ($\abs{i-j}=2$) as $J_{t_i}$. In particular, $J_{v_i}$ and $J_{w_i}$ express tunneling through the odd and even barriers of the superlattice potential respectively. Finally, we consider $h_{i,j}$ = 0 for $\abs{i-j} > 2$ because these terms are significantly suppressed, due to the fact that they correspond to tunnelling through at least three consecutive barriers. 

In the case of periodic boundary conditions (PBC), as the potential landscape is also periodic, it is evident that $\mu_i=\mu,~~J_{v_i}=J_v,~~J_{w_i}=J_w$ and $J_{t_i}=J_t~~\forall i$. In essence, all the corresponding sites have the same potential environment. However, we are interested in the realization of edge states via a finite optical lattice so the latter has to possess both a left and a right end at a finite $x$-value. This necessarily implies that the external potential has to be confining, breaking the translation symmetry. In our analysis the confinement comes from the hard wall boundary conditions (HWBC) or from the boundary extension described above.
\begin{figure}[ht]
\begin{center}
\begin{overpic}[scale = 0.57]{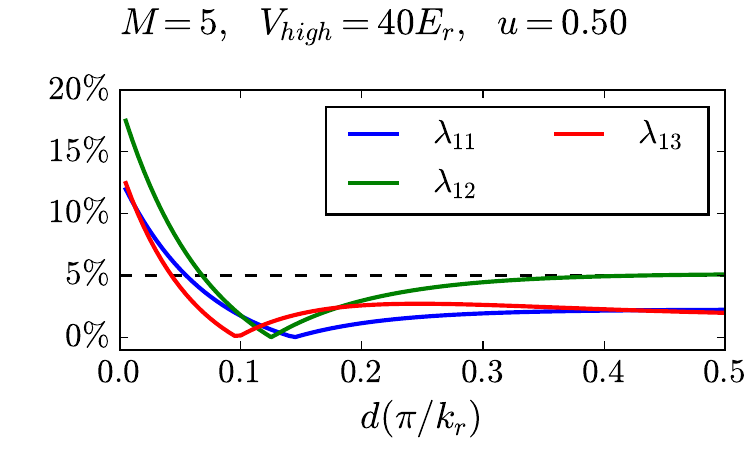}
\put(8,55.75) {\textbf{(a)}}
\end{overpic}
\qquad
\begin{overpic}[scale = 0.57]{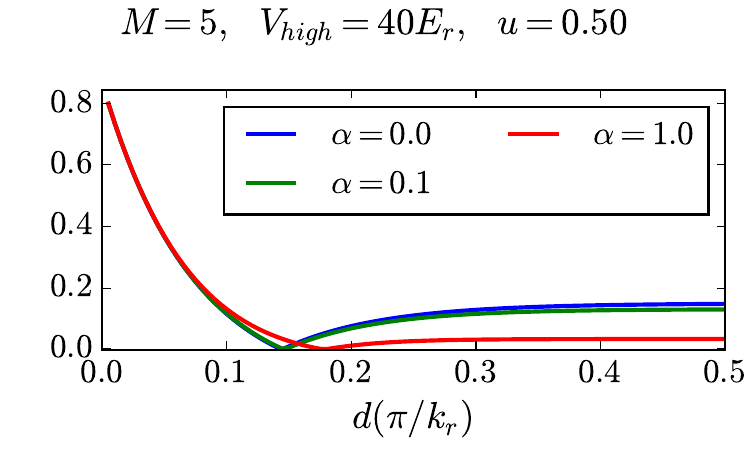}
\put(-2,22) {\rotatebox{90}{$\rho\left(\delta \mathcal{H}\right)$}}
\put(8,55.75) {\textbf{(b)}}
\end{overpic}
\end{center}
\caption{(a) The elements $\lambda_{11}, \lambda_{12}$ and $\lambda_{13}$ versus the length of the linearly extended boundary $d$ and $\alpha = 0$. (b) The spectral radius of the matrix $\delta \mathcal{H}$ versus the length of the linearly extended boundary $d$ and three cases of the slope $\alpha = 0,0.1,1$ in units of $E_r\cdot k_r$.}
\label{fig:Spectral_Radius}
\end{figure}
Our mapping is mostly affected by the boundaries in the case of the topologically non-trivial configuration of the potential landscape. In particular, the tails of the localized states related to the edge sites are forced to zero in a sharp (unnatural) way. Another perspective of the finite boundaries effect, is that every site of the discrete system is mapped to a minimum of the superlattice potential that features a different environment. With that in mind, we define:
\begin{equation}
h_{i,j}=
\begin{cases}
\mu+ \delta \mu_i   & i = j \\
J_{v,w} + \delta J_{v_i,w_i} & \abs{i - j} = 1 \\
J_t+ \delta J_{t_i} & \abs{i-j} = 2 \\
\end{cases}
\end{equation}
where $\mu$, $J_{v}$, $J_{w}$ and $J_t$ are the bulk values corresponding to the center sites ($M$ and $M\pm1$, $M\pm2$), while $\delta \mu_i,~\delta J_{v_i},~\delta J_{w_i}$ and $\delta J_{t_i}$ denote the deviations from these bulk values at each site. Those deviations will be present in any confining potential, since they express the gradual modification of the potential environment towards the edges of the system. So, even though the following analysis is based on the consideration of HWBC or of a linear boundary, the techniques that we develop here can be applied to any kind of confining potential. The Hamiltonian of the system can be written as:
\begin{equation}
\mathcal{H}_{TB} = \mathcal{H}^{(0)} + \delta \mathcal{H}
\end{equation}
where $\mathcal{H}^{(0)}$ has as elements the bulk values, and $\delta \mathcal{H}$ the deviations. 
\begin{figure*}[ht!]
\begin{center}
\begin{overpic}[width=0.98\textwidth]{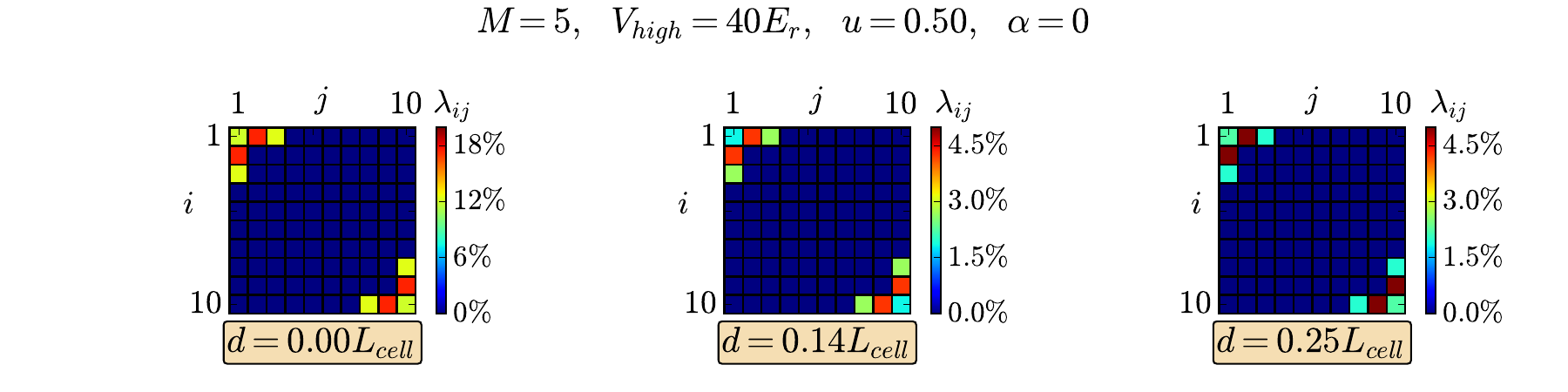}
\put(27,23.25) {\textbf{(a)}}
\end{overpic} \\[25pt]
\begin{overpic}[width=0.98\textwidth]{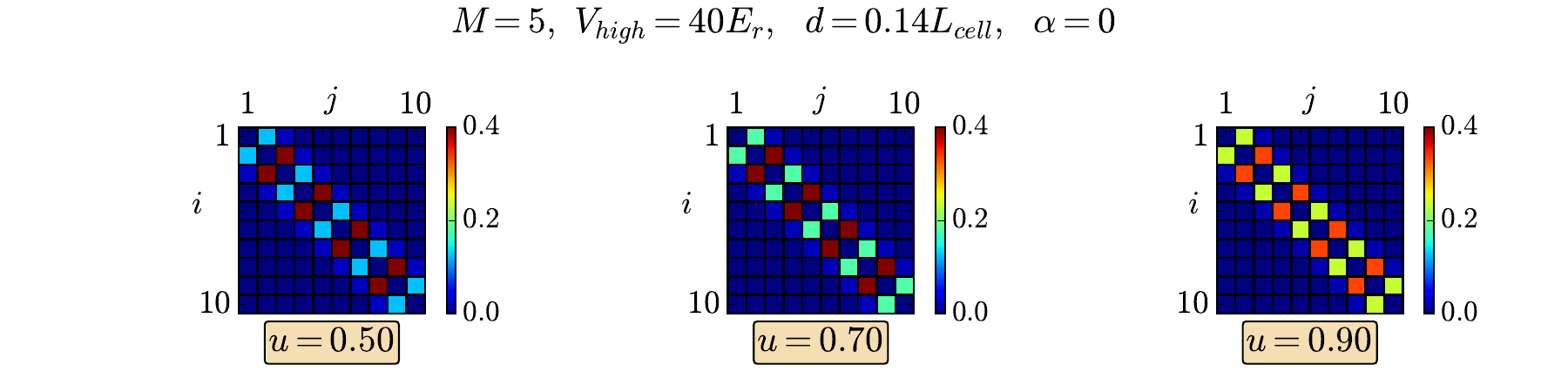}
\put(25.5,23.25) {\textbf{(b)}}
\put(28,19.25) {$\mathcal{H}^{TB}_{ij}$}
\put(59.5,19.25) {$\mathcal{H}^{TB}_{ij}$}
\put(90.5,19.25) {$\mathcal{H}^{TB}_{ij}$}
\end{overpic}
\end{center}
\caption{(a) The $\lambda$ matrix for a system of $M=5$ cells, $V_{high} = 40E_R$ and $u = 0.5$ for different values of the length $d$ of the linear extension of the potential area ($\alpha = 0$). (b) The $\mathcal{H}_{TB}$ matrix for a system of $M=5$ cells, $V_{high} = 40E_R$, $d = 0.14L_{cell}$ (and $\alpha = 0$) for different values of the ratio $u$ of the heights of the potential. In all cases, we have used color plots as a visual tool to display the elements of the matrices. The elements $\mathcal{H}^{TB}_{ij}$ are in units of $E_r$.}
\label{fig:VisualTool}
\end{figure*}
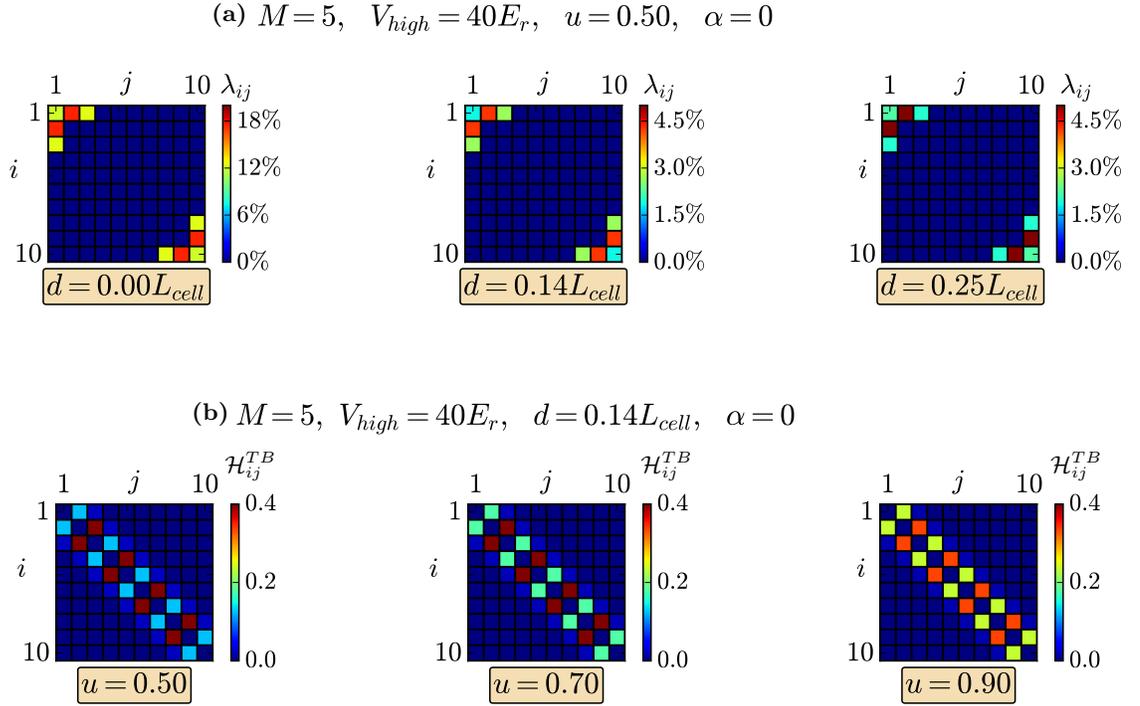
\subsection{Minimising the $\vb*{\delta \mathcal{H}}$ term}
\label{delta_H_zero}
In order to examine the behaviour of the $\delta \mathcal{H}$ term, we can define the $\lambda$ matrix, with elements:
\begin{equation}
\lambda_{ij}=\big|\delta \mathcal{H}_{ij}\big|/\big|\mathcal{H}^{(0)}_{ij}\big|
\end{equation}
expressing the relative deviation from the bulk values. As we can see in Fig. \ref{fig:VisualTool}(a), the strongest (and only significant) deviations arise from the terms that correspond to the endpoints, in accordance with the results of the continuous system. Furthermore, we see that the elements of the $\lambda$ matrix depend on the parameters of the linear extension of the potential domain. Specifically, as we can see in Fig. \ref{fig:Spectral_Radius} (a), for fixed $\alpha$ and increasing $d$, they decrease down to certain minimum values and then increase until they stabilize and become $d$-independent. In order to determine the critical length $d_{crit}$ for which the elements of the $\lambda$ matrix are minimized, we employ the spectral radius \cite{Horn_Johnson_1985}. 
Namely, we solve the eigenvalue problem of the matrix 
\begin{figure}[ht]
\begin{center}
\begin{overpic}[scale = 0.63]{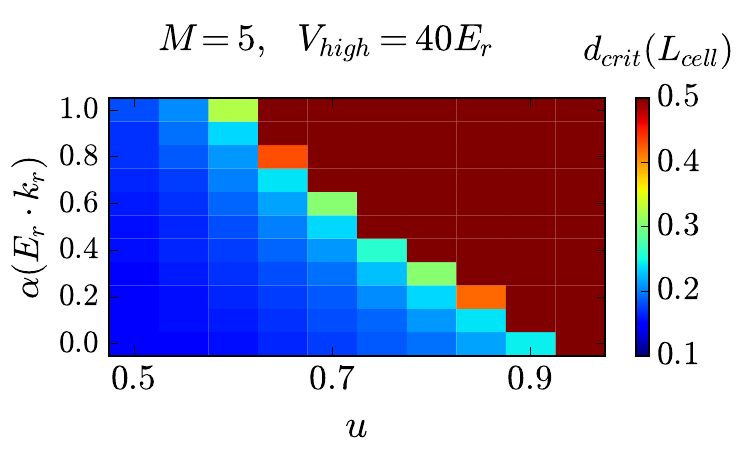}
\end{overpic}
\end{center}
\caption{Color plot of the $d_{crit}$ in units of $L_{cell}$ defined as the point of minimum of the spectral radius versus the ratio of the heights of the potential barriers $u$ and the slope $\alpha$. The area colored in deep red indicates the parameter region where the spectral radius has no minimum.}
\label{fig:d_crit}
\end{figure}
$\delta\mathcal{H}$ for specific values of the parameters $d$ and $\alpha$, and we find the eigenvalue with the maximum absolute value, which is the spectral radius $\rho(\delta\mathcal{H}$) of the matrix. We repeat this process for different values of the parameter $d$. Finally, we identify as $d_{crit}$ the value of $d$ for which $\rho(\delta\mathcal{H})$ is minimized. As shown in Fig.~\ref{fig:Spectral_Radius} (b) for three different cases of the slope $\alpha$ there is indeed a $d_{crit}$. Moreover, we observe in Fig. \ref{fig:d_crit} that for specific values of the parameters of the system there is no minimum in the spectral radius, and one should find a different way to define $d_{crit}$. For example, in those cases $d_{crit}$ could be defined as the value of $d$ for which the elements of the $\lambda$ matrix become $d$-independent.

\subsection{The behaviour of the $\vb*{\mathcal{H}_\text{NNN}}$ term}
\label{NNN_term}
\begin{figure*}[ht]
\begin{center}
\begin{overpic}[scale = 0.35]{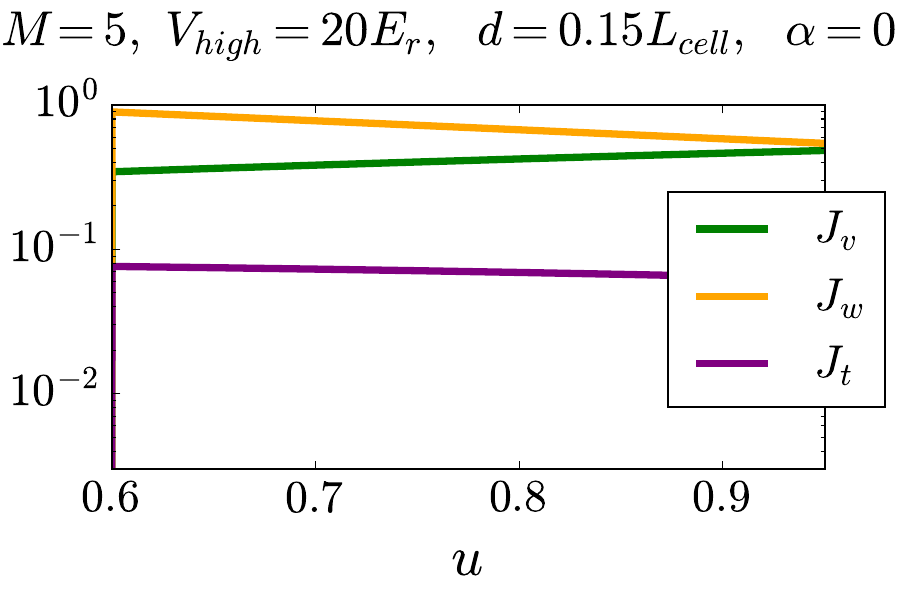}
\put(14,30) {\textbf{(a)}}
\end{overpic}~~~~
\begin{overpic}[scale = 0.35]{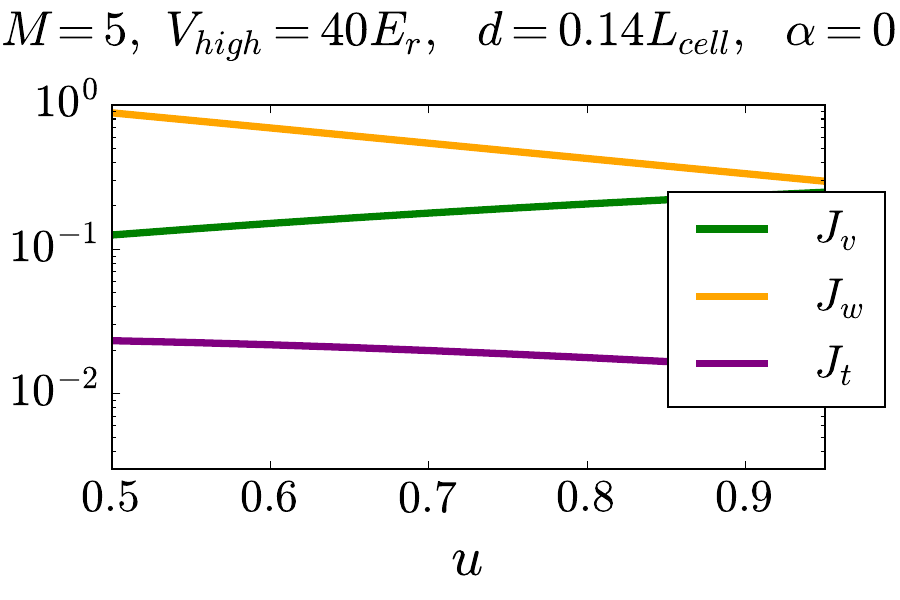}
\put(14,31.5) {\textbf{(b)}}
\end{overpic}~~~~
\begin{overpic}[scale = 0.35]{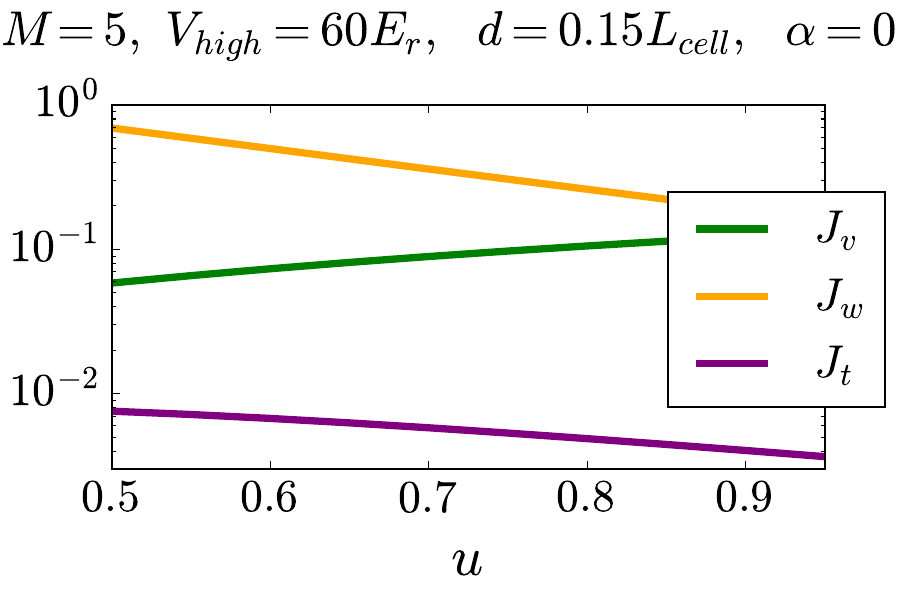}
\put(14,29) {\textbf{(c)}}
\end{overpic}
\end{center}
\caption{The $J_v,J_w$ and $J_t$ hopping amplitudes values in units of $E_r$ for systems with $M=5$, 
(a) $V_{high}=20E_r$ and $d=0.15L_{cell}$  (b) $V_{high}=40E_r$ and $d=0.14L_{cell}$ and
(c) $V_{high}=60E_r$ and $d=0.15L_{cell}$. In (a) the x-axis starts from $0.6$, since for $u<0.6$ not all eigenergies of the first band lies below the lowest barrier ($V_{low}$).}
\label{fig:J_s_values}
\end{figure*}

In this subsection we consider systems with $d=d_{crit}$ so that we can omit the $\delta \mathcal{H}$ term. Hence, we can write $\mathcal{H}_{TB}$ as:
\begin{equation}
\mathcal{H}_{TB} = \mathcal{H}^{(0)} =  \mathcal{H}_{\mu} + \mathcal{H}_{SSH} + \mathcal{H}_\text{NNN} 
\end{equation}
where $\mathcal{H}_{\mu}$ and $\mathcal{H}_\text{NNN}$ express the on-site and next-nearest neighbour hopping terms respectively. We can also neglect the $\mathcal{H}_\mu$ term, since it represents an overall offset in the energy scale that does not affect the phenomenology of our systems. So we reach to $\mathcal{H}_{TB} = \mathcal{H}_\text{SSH} + \mathcal{H}_\text{NNN}$ which is clearly the exact same as Eq. \eqref{eSSH_eq}. Thus, we have ultimately obtained the usual SSH model with the addition of NNN hopping terms (eSSH). In Fig. \ref{fig:VisualTool} (b) we present how the final form of the TB Hamiltonian is affected by the ratio of the heights of the two potential barriers $u$. For fixed $V_{high}$ and increasing $u$ we see $J_v$ increasing, $J_w$ decreasing and $J_t$ slightly decreasing (see also Fig. \ref{fig:J_s_values} for the exact values for three different cases of $V_{high}$). 

\begin{figure}[ht]
\begin{center}
\begin{overpic}[scale = 0.5]{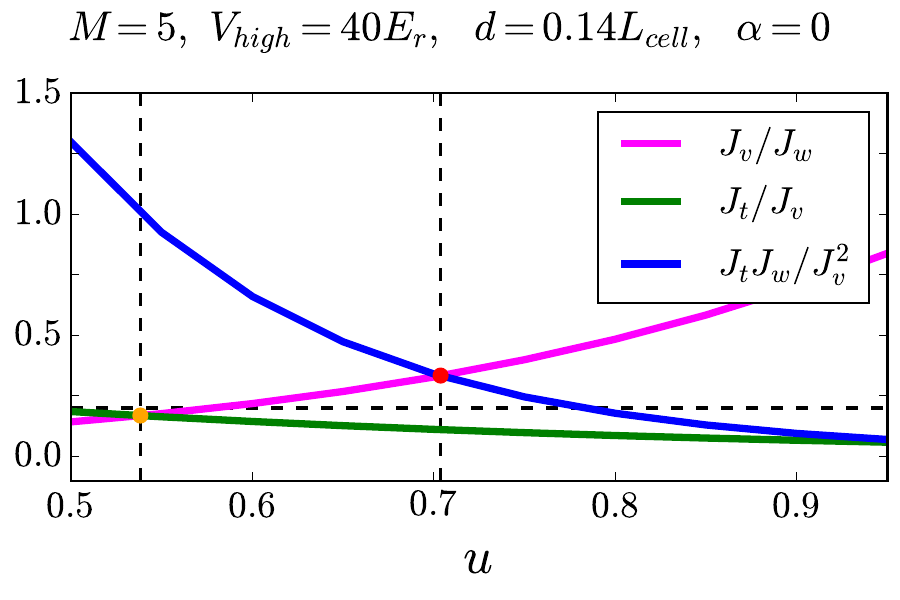}
\end{overpic}
\end{center}
\caption{The ratios $J_v/J_w$, $J_t/J_v$ and $(J_t/J_v)/(J_v/J_w)$ for the hopping amplitudes of a system with $M=5$, $V_{high}=40E_r$ and $d=d_{crit}$. The horizontal dashed line shows the $0.2$ ratio value, while the vertical dashed lines show the values of $u$ where the crossing of $J_v/J_w$ with $J_t/J_v$ and $(J_t/J_v)/(J_v/J_w)$ occurs.}
\label{fig:J_s_ratios}
\end{figure}

In order to neglect the $\mathcal{H}_\text{NNN}$ term we have to make a strong assumption, since we have to  take into account the different scales of magnitude of the hopping amplitudes. To this end, we define the ratios that we present in Fig. \ref{fig:J_s_ratios}. Specifically, $J_v/J_w$ and $J_t/J_v$ express relative sizes of the weak to strong and NNN to weak hopping amplitudes, respectively. As we increase $u$, we observe $J_v/J_w$ increasing and $J_t/J_v$ decreasing. Therefore, there is a specific value of $u$ for which the two ratios are crossing (orange point in the graph of Fig. \ref{fig:J_s_ratios}). We could in principle define this value as a $u_{crit}$ and omit the $\mathcal{H}_\text{NNN}$ term for $u>u_{crit}$. Alternatively, we could define $u_{crit}$ as the value of $u$ for which $J_v/J_w$ and $(J_t/J_v)/(J_v/J_w)$ are crossing (red point in the graph of Fig. \ref{fig:J_s_ratios}). The ratio $(J_t/J_v)/(J_v/J_w)$ expresses how $J_v$ varies from small (close to $J_t$) to large (close to $J_w$) orders of magnitude (for fixed $V_{high}$ and increasing $u$). This is naturally a more strict definition, since for the first definition we omit the $\mathcal{H}_{NNN}$ term when $J_v/J_w>J_t/J_v$ while for the second when $J_v/J_w>\sqrt{J_t/J_v}$. We mention that for both definitions $u_{crit}$ is $V_{high}$ dependent. 

We close this subsection by noting that both SSH and eSSH support edge states. However, when it comes to their experimental realization via an optical lattice, the NNN terms will always be present if both wells are not deep enough. Thus, the eSSH model is the easier to be implemented.
We note that next-neighbor tunneling terms in the SL were,
 e.g., considered in the context of Thouless pumps, where they can lead to a deviation of the pumped charge \cite{PhysRevA.104.063315}.

\subsection{Experimental Considerations}

An extension of the boundary as proposed here is experimentally feasible by combining an optical lattice with an arbitrary optical dipole potential projected via a high resolution imaging system. This allows shaping the potential at the scale of a single lattice site and also to prepare desired lattice occupations \cite{Weitenberg_2011, Preiss_2015} and was recently used to create sufficiently steep walls for the creation of edge states in 2D systems \cite{braun2023realspace, yao2023observation}. The above analysis shows that the correct mid-gap states are reached for a range of parameters without fine tuning, making it experimentally feasible. The selective preparation of the edge states could be achieved, e.g., by an appropriate charge pump or by heating away all other atoms via an amplitude modulation that leaves the edge state unaffected.

\section{Summary and Perspectives}\label{sec:SummaryAndOutlook}

We have studied the mapping of a finite continuous optical lattice, within the tight-binding approximation, to an SSH and an eSSH model. We revealed the complications that emerge from the confinement of the system, when hard wall boundary conditions are considered. Specifically, we found that especially in the topologically non-trivial case, there is a substantial discrepancy in the characteristics of the edge states of the system between what is expected for the SSH model and the results provided by the ED. We found that this is related to the fact that the potential environment is gradually changing from the bulk towards the ends of the system, due to the confinement. We then proceeded by providing a solution to this problem through a linear extension of the potential area. In detail, we established and used qualitative criteria in order to describe the minimization of the effects related to the HWBC. We also examined the behaviour of the NNN hopping terms in system with experimentally feasible depths of wells.

The goal of the present work was to reveal the problems occurring in the mapping of a continuous optical lattice potential to a discrete SSH model when finite size and open boundary conditions are considered. From a theoretical point of view, these problems are resolved with the boundaries extension we proposed here. Nevertheless, for the design of an optimal experimental set-up capable to detect the associated edge states, a more advanced optimization procedure, involving a more general functional form of the potential extension, may be needed. This functional form should be eventually also adapted to the experimental requirements for the specific set-up. Additionally, the observability of the continuous formed edge states may be influenced significantly by their dynamics. However, a corresponding dynamics analysis goes beyond the scope of the present work and will be investigated in the future. 

The exact realization of tight-binding models for ultracold atoms was recently also considered for tweezer arrays, including the homogeneity of the on-site interaction strength \cite{Wei_2024}. The study of interacting superlattice systems is particularly interesting, because the bosonic model with weak interactions lacks chiral symmetry and the edge states are no longer mid-gap \cite{PhysRevLett.110.260405}.
Further interesting extensions would be to consider optimal parameters for mid-gap states in the bulk of the system induced by impurities \cite{Lang_2014} or by a step in the confining potential \cite{PhysRevLett.110.260405}, as well as optimal boundaries when combining superlattices with a Floquet drive to realize the AIII symmetry class \cite{PhysRevLett.119.115301}.
It would also be interesting to perform a similar anaylsis for engineered systems with indirect band gaps, where surprisingly no localized edge states were found \cite{PhysRevResearch.1.033197}.

\section*{Acknowledgments} 

This work (P.S. and C.W.) has been
supported by the Cluster of Excellence “Advanced Imaging of Matter” of the
Deutsche Forschungsgemeinschaft
(DFG)-EXC 2056, Project ID No. 390715994. \\[15pt]
\newpage
\section*{APPENDIX}
\appendix
\section{Edge localized states of very large/deep lattice}
\label{app:LargeSystem}
\begin{figure}[ht!]
\begin{center}
\begin{overpic}[scale = 0.425]{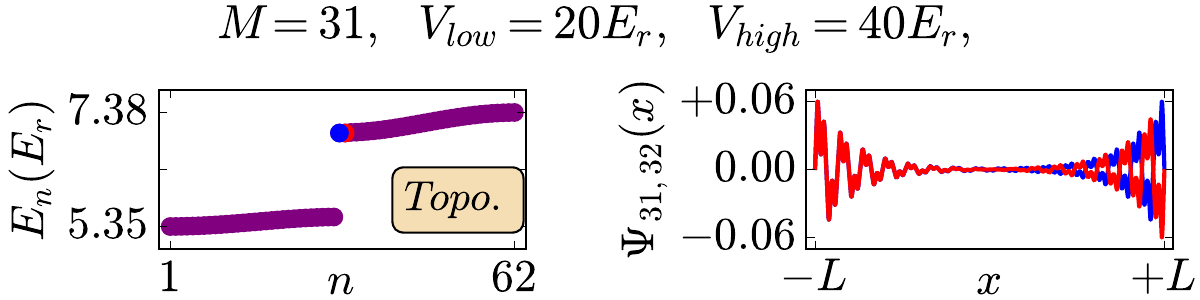}
\put(12,22.25) {\textbf{(a)}}
\end{overpic}
\\[10pt]
\begin{overpic}[scale = 0.425]{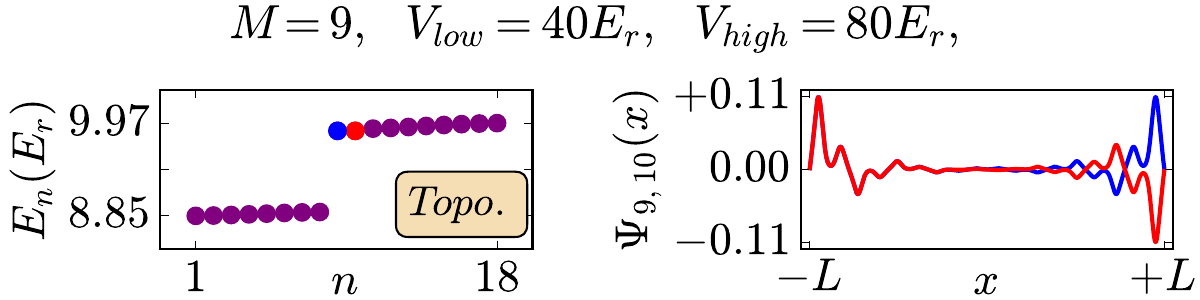}
\put(12,22.25) {\textbf{(b)}}
\end{overpic}
\end{center}
\caption{The energy spectrum and the profiles of the corresponding edge localized states for a very (a) large and (b) deep lattice ($x$ is in units of $k_r^{-1}$).}
\label{fig:LargeDeep}
\end{figure}

In Fig. \ref{fig:LargeDeep} we present the energy spectrum and the edge localized states of a very (a) large and (b) deep lattice when HWBC are considered, i.e. without the extension of the potential domain. In both cases the energies of the two edge localized states feature an energy shift towards the upper band. This is to be expected, since it is related to the consideration of HWBC for the system. However, in the case of the very deep lattice we observe that the profiles of the two edge localized states look similar to those of an eSSH model. 
This is remarkable, since the discrepancy on the energies is still present.

\bibliographystyle{apsrev4-1}
\bibliography{ref_drops}

\end{document}